# Adherend surface roughness effect on the mechanical response of adhesive joints

Hamed Zarei[1*], Maria Rosaria Marulli [1], Marco Paggi[1], Riccardo Pietrogrande[2], Christoph Üffing[3], and Philipp Weißgraeber[3]

[1]IMT School for Advanced Studies Lucca, Piazza San Francesco 19, 55100 Lucca, Italy

[2]Automotive Electronics, Engineering Technology Polymers (AE/ETP), Robert Bosch GmbH, Renningen, Germany

[3]Corporate Research and Advance Engineering (CR), Robert Bosch GmbH, Renningen, Germany

**Abstract**

The present contribution focuses on the effect of adherend surface roughness on the strength of adhesive joints, which are particularly cost-effective and extensively applied in a wide range of industrial applications. However, the reliability of such solutions is a critical concern for the integrity of commercial products. To gain a deeper understanding on the effect of roughness, an extensive experimental campaign is proposed, where thermoplastic substrates are produced with a specified roughness, whose characterization has been performed using a confocal profilometer. Elastic strips are then bonded onto such substrates using Silicone adhesive while controlling the adhesive thickness. Peeling tests are finally carried out and the effects of joint parameters such as surface roughness, adhesive thickness, and loading rate are discussed in detail. Eventually, it is demonstrated that the surface roughness can increase the adhesion energy of joints depending on the value of a ratio between the adhesive thickness and the root mean square elevation of roughness.

[*]Corresponding author. E-mail: hamed.zarei@imtlucca.it

| Nomenclature | | | |
|---|---|---|---|
| α | Adhesive thickness to surface roughness ratio | $R_z$ (μm) | Average roughness |
| Δ (mm) | Peel extension | t (mm) | Adhesive thickness |
| θ | Peeling angle | w (mm) | Strip width |
| σ (μm) | Root mean square of the height filed | z (μm) | Height field |
| F (N) | Peeling force | $\bar{z}$ (μm) | Average of height field |
| G (N/mm) | Adhesion energy | | |

## 1. Introduction

Design for Manufacturing and Assembly (DFMA) approach, which concerns the assembling procedure of components, is one of the most important stages in engineering design. In this area, there are several techniques to attach subparts to each other, including adhesives, especially for bonding dissimilar materials. Adhesives are widely used materials in both engineering and biology industries [1–4] and the enhancement of interfacial properties is one of the key research topics [5–7].

The measurement of the interfacial mechanical properties between adhesive and adherend, which is essential to assess an optimized superior interface resistance, can be performed through peeling tests. Peeling of an adhesive layer from a substrate is an energy-driven process. Therefore, peeling resistance describes the evaluation of the effective bond strength and of the adhesive fracture energy, which is the energy needed to create a new interfacial area as the peel arm is pulled away from the substrate. Many researchers in a variety of scientific areas have studied the mechanics of peeling for numerous materials, either with theoretical and numerical models or experimentally [8–12]. Based on the achievements reported in the literature, a wide quantity of parameters can influence the mechanical response of the adhesives. Overall, there are three main groups that could be used for parameters' classification: (i) the intrinsic adhesive properties which include chemicals, molecular structure, strength, viscoelasticity, etc. [13–16]; (ii) the substrate interaction involving its compatibility with the adhesive, the adhesive application process, surface morphology, etc. [17–19]; (iii) environmental factors that can degrade bonding over time, such as temperature, moisture, etc. [20–24]. To design an optimized adhesive, all the above-mentioned factors should be taken into account.

The present study deals with the impact of surface roughness on the strength of adhesive joints. It is noteworthy that, to the best of the authors' knowledge, hitherto there is no published research concerning the effect of prescribed surface roughness on the peeling response of adhesive joints. To investigate such a roughness effect, we laid out an extensive experimental campaign by producing substrates with different prescribed surface roughnesses. Subsequently, after bonding flexible but inextensible strips onto such substrates, the mechanical response of the prepared samples was assessed through peeling tests.

The present article is organized as follows: through the next section, the whole experimental scenario including substrates fabrication, surface roughness characterization, bonding strip onto the substrate, and lastly the peeling test setup are addressed. In Section 3, the results of the peeling tests and the impact of some parameters such as loading rate effect, adhesive thickness and surface roughness on the adhesive strength is discussed. Eventually, the main conclusions are given in Section 4.

## 2. Experimental methods
### 2.1. Substrate fabrication

To investigate the roughness effect on the strength of adhesive joints, two specimen types are designed: (i) smooth substrate, and (ii) substrate including a final rough part on its surface. Polybutylene Terephthalate (PBT) polymer-based substrates are fabricated through injection molding. For specimens with roughness, steel inserts with specified roughness were employed to generate a rough area on those samples after injection molding (see Figure 1). The rough steel inserts were such that they could transfer their roughness onto the plastic substrate. Different rough inserts were used, with $R_z$ ranging from 0.6 to 7.5 µm. The total specimen size is 120×25×2 mm³, with 25×25 mm² as a final rough part. Figure 2 shows all specimens that are considered in the present study.

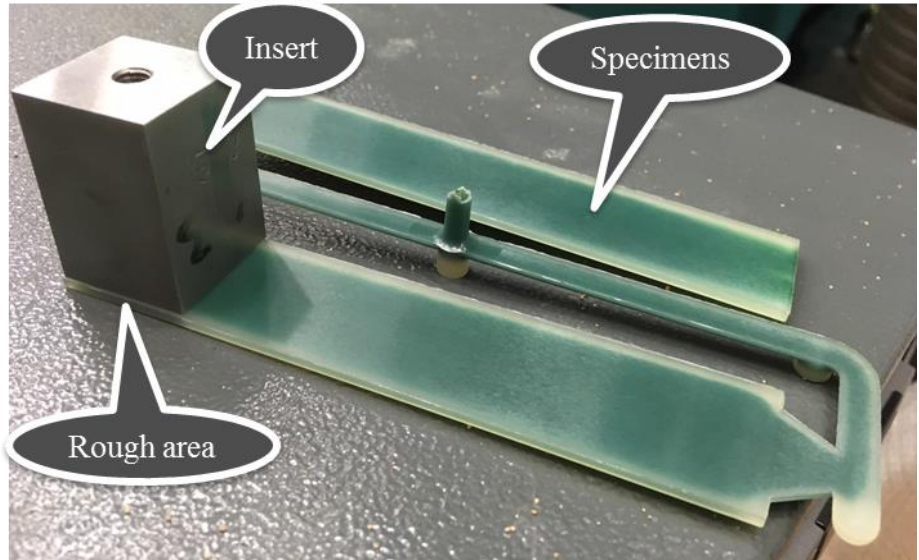

*Figure 1. Making the specified roughness using inserts.*

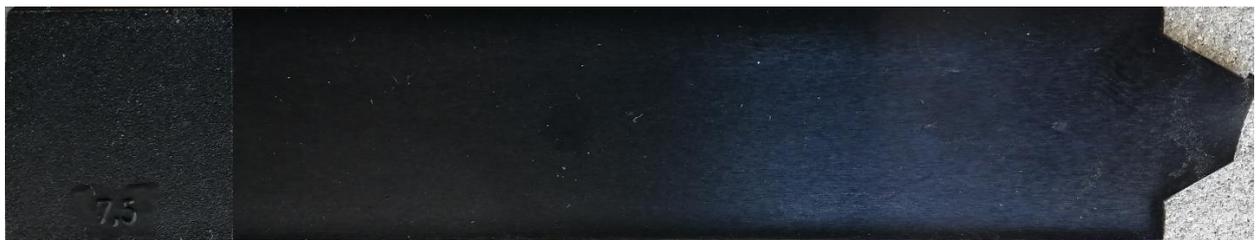

a) Whole substrate with rough (left end) and smooth (on its right) portion.

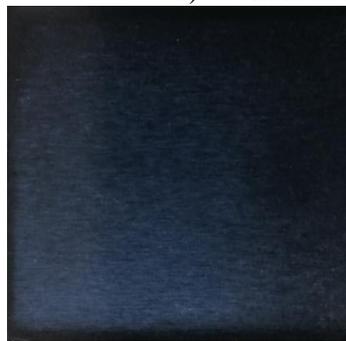 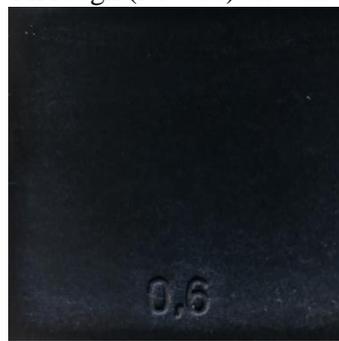 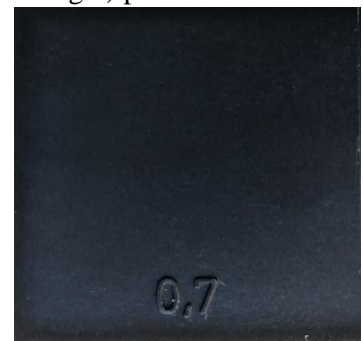

Smooth  —  $R_z$=0.6 µm  —  $R_z$=0.7 µm

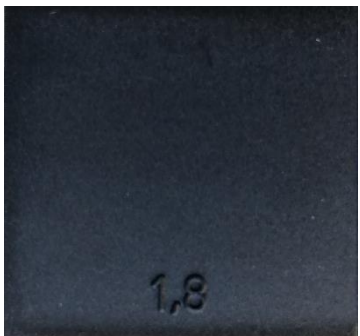 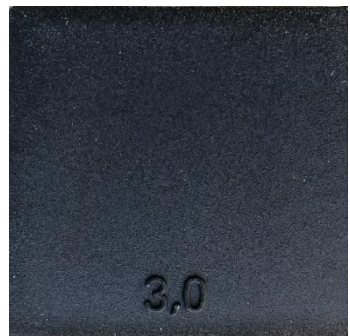 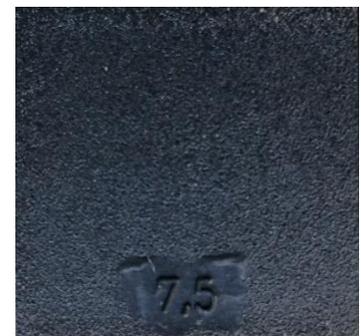

$R_z$=1.8 µm  —  $R_z$=3.0 µm  —  $R_z$=7.5 µm

b) Detail on the rough part

*Figure 2. Plastic substrates obtained by means of rough steel inserts with different $R_z$ values.*

**2.2. Roughness characterization**

The surface roughness obtained on the samples might be different from the metallic insert one, since it results from the injection molding process. In order to achieve a reliable characterization of the roughness transferred to the samples, the rough surfaces have been acquired using the non-contact confocal profilometer LEICA DCM3D available in the MUSAM-Lab at the IMT School for Advanced Studies Lucca. It is equipped with different lenses in order to provide surface scans with different magnification (10x, 20x and 100x), see Figure 3. For the present study, the fine level of roughness chosen required the use of the highest resolution. Additionally, the stitching option has been used to scan an extended topography of progressive increasing size, up to which the scanned sample could be considered as statistically representative for the whole surface. Such topographies for the finest ($R_z$= 0.6 μm) and coarsest surfaces ($R_z$= 7.5 μm) are shown in Figure 4 and Figure 5, respectively. Moreover, properties of the extended topographies of a scanned surface with up to 8×8 stitching samples are summarized in Table 1.

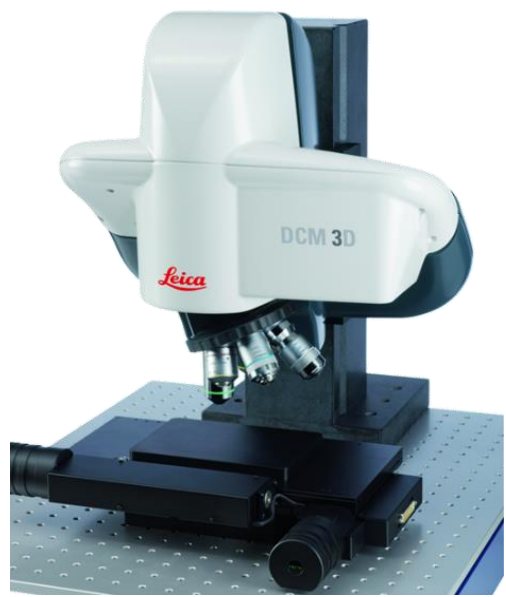

*Figure 3. Non-contact confocal profilometer (Leica DCM 3D).*

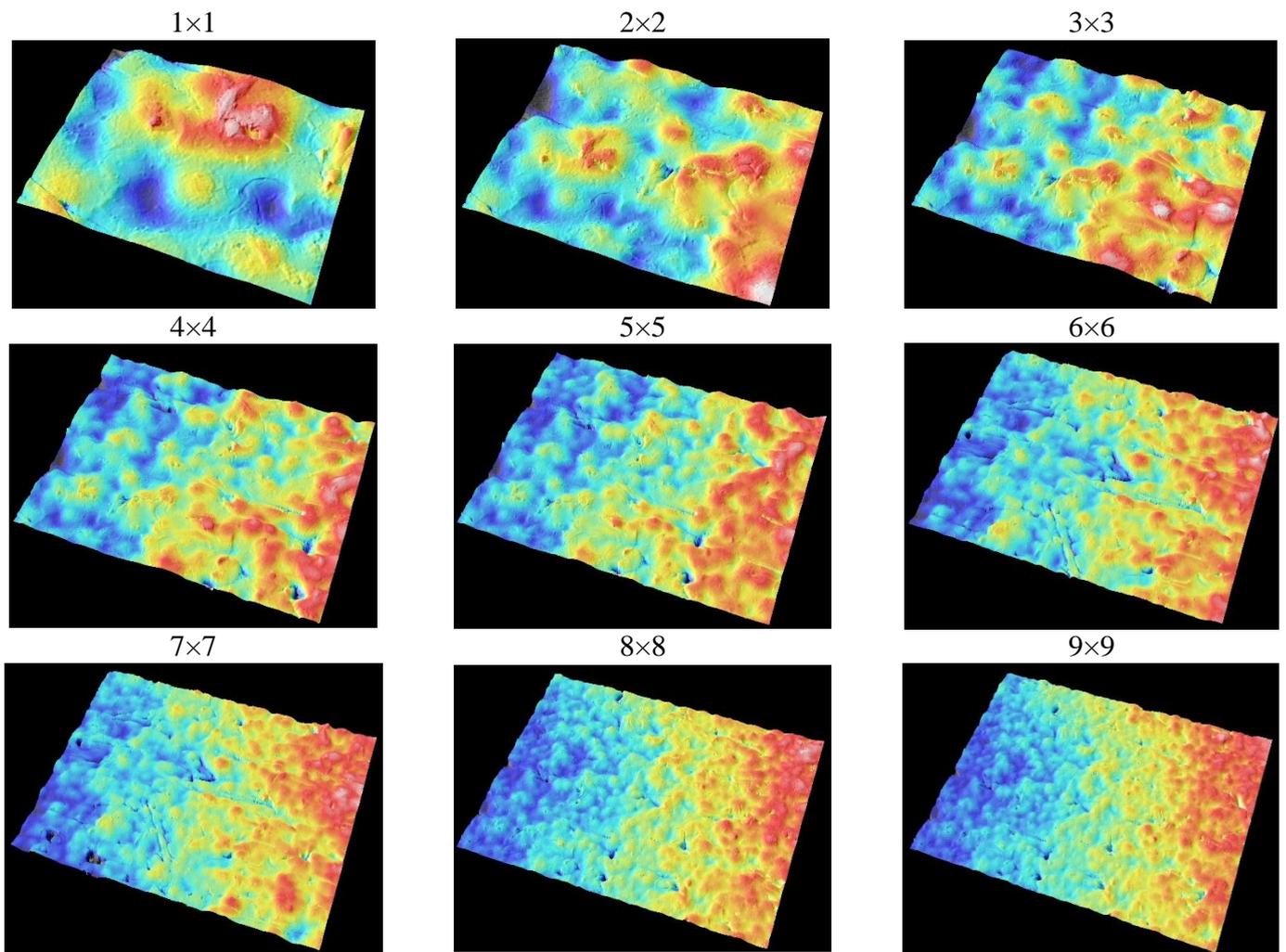

*Figure 4. Scanned extended topography of substrate with the roughness of $R_z$=0.6 µm (elevation range: ±8.0 µm), while increasing its size through n x n stitching of individual scans.*

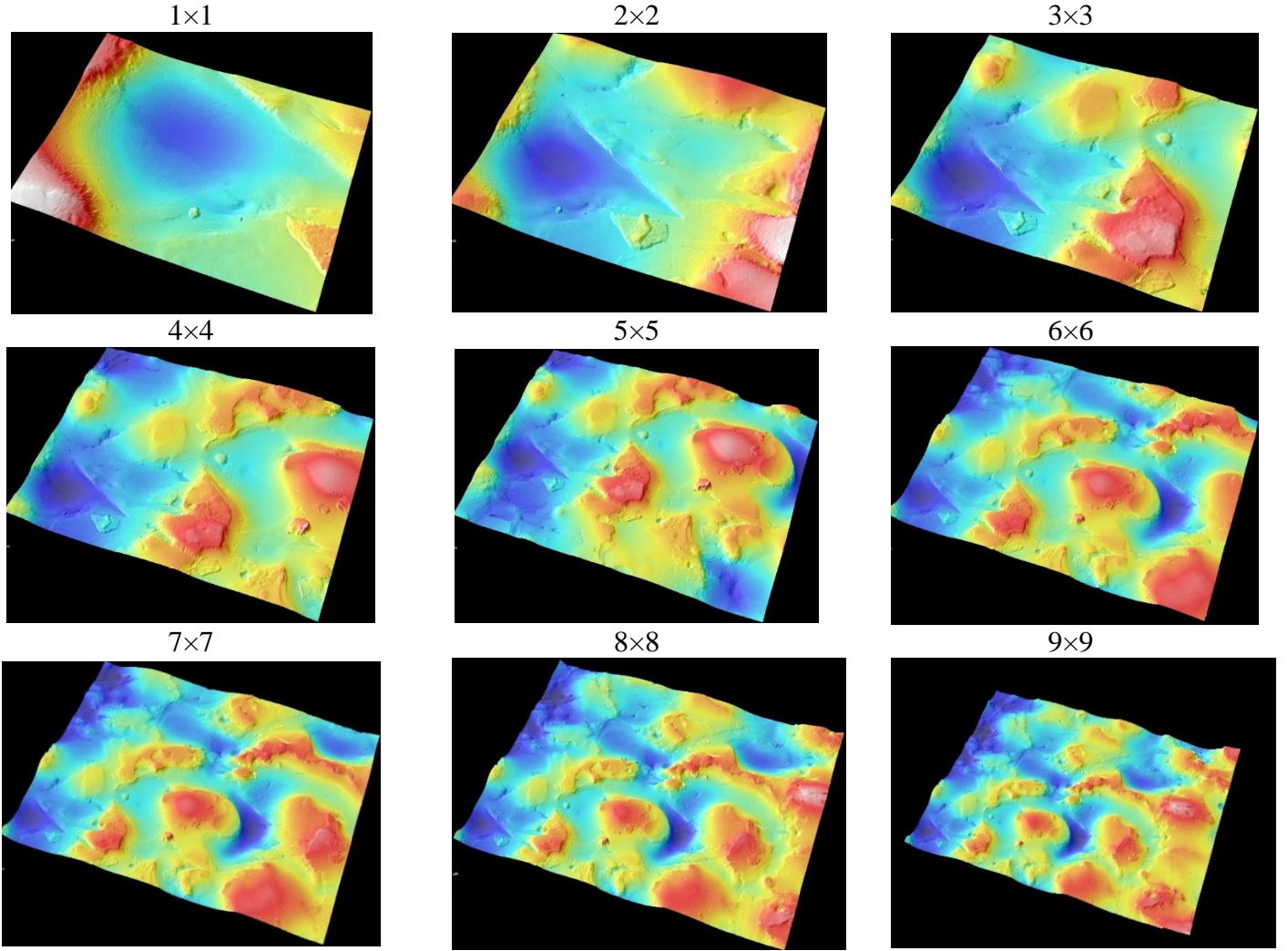

*Figure 5. Scanned extended topography of substrate with the roughness of $R_z$=7.5 μm (elevation range: ±38.0 μm) while increasing its size through n x n stitching of individual scans.*

*Table 1. Extended topography properties of a scanned surface with different grid sizes.*

| Grid Size | 1×1 | 2×2 | 3×3 | 4×4 |
|---|---|---|---|---|
| Area (μm$^2$) | 127×95 | 229×172 | 332×249 | 434×325 |
| Number of heights | 768×576 | 1382×1036 | 1996×1496 | 2610×1956 |
| **Grid Size** | **5×5** | **6×6** | **7×7** | **8×8** |
| Area (μm$^2$) | 536×402 | 638×478 | 740×555 | 843×632 |
| Number of heights | 3224×2416 | 3838×2876 | 4452×3336 | 2533×1898 |

As the next step, we have quantified the minimum size of the surface to be statistically representative of roughness. This aspect is crucial, since the number of distinct features increases while adding more samples areas using stitching. After a certain size, we expect to include enough valleys and peaks such that the overall ensemble is able to fully characterize the statistical features of the overall surface. To the best authors' knowledge, no similar in-depth studies on the determination of the statistically representative surface size from real profilometric data are available in the literature. The problem has been herein addressed by exploiting the Dynamic Space Warping (DSW) algorithm [25] using as input the normalized distributions of the experimentally acquired surfaces' height fields. Figure 6 shows the probability distributions of the normalized height field for different stitching numbers for the substrate realized by molding with steel indenters having $R_z$= 7.5 μm. By increasing the scanned area, the experimental probability density functions tend to converge as expected, so that for surfaces larger than 8×8, they qualitatively coincide, which means that the observed portion of the surface becomes large enough to be statistically representative of the whole sample. However, DSW can provide a quantitative indicator to assess the distance (mismatch) between two probability distribution curves, representing the similarity between them. Based on such a distance measure obtained from the DSW algorithm (Figure 7), the similarity is very high for stitched surfaces having more than 8×8 samples, with a progressive tendency not to vary anymore by further increasing the surface size. Therefore, the surface obtained by stitching 8×8 samples can be considered as representative surface for the one produced by the steel insert with $R_z$= 7.5 μm. The representative surface sizes corresponding to all the other roughness values are presented in Table 2.

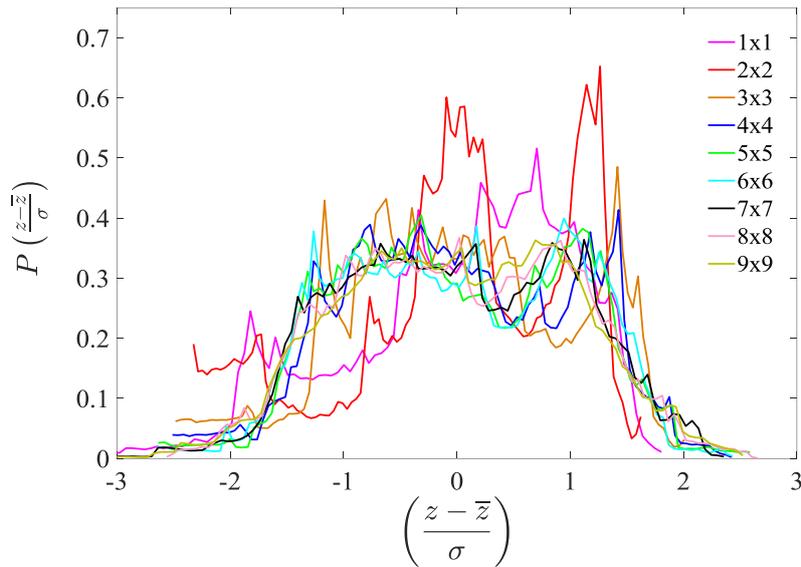

*Figure 6. Normalized distribution of the height field for different stitching number.*

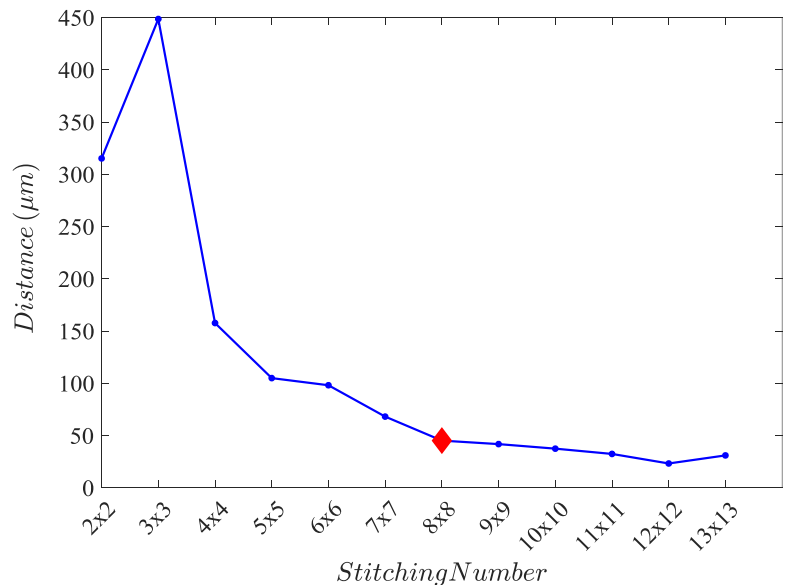

*Figure 7. Similarity value obtained from the DSW algorithm vs. size of the stitched surface, for the substrate created through the steel insert with $R_z$=7.5 μm.*

Table 2. The representative surface size of rough surfaces generated through the steel insert with different $R_z$.

| $R_z$ (μm) | Minimum surface size based on stitching |
|---|---|
| 7.5 | 8×8 |
| 3.0 | 7×7 |
| 1.8 | 6×6 |
| 0.7 | 6×6 |
| 0.6 | 5×5 |

### 2.3. Sample preparation

For peeling test, a two-component Silicone-based adhesive is used to bond HELIOX PV FERON NEOX CPC 300 onto the substrate. The HELIOX strip was tested in the MUSAM-Lab to characterize under tensile loading. It was chosen since it is a stiff material with Young's modulus of 4.9 GPa and a tensile strength of 180 MPa. These values assure its linear elastic response for the entire peeling test. It should be noticed that HELIOX PV with a 5 mm width has been used as a flexible and inextensible strip.

In order to produce the joined specimens, a specific fixture capable of controlling the adhesive thickness was designed and manufactured within the project (Figure 8). Firstly, the substrate is fixed in the mold, and then the clamp at the upper part of the fixture is used to keep the strip in the correct position during the adhesive feeding. By dispensing the adhesive onto the substrate, laying down the strip on top of it and manually sweeping the squeegee with a constant rate, the bonding of the strip onto the substrate is achieved with a good reproducibility. It is noteworthy that the squeegee is allowed to move vertically as well, and its position can be regulated by tightening the two embedded screws. Therefore, the desired vertical position is ensured by using a metal index with a specific thickness that will correspond to the final adhesive thickness. An example of prepared sample for the peeling test is shown in Figure 9.

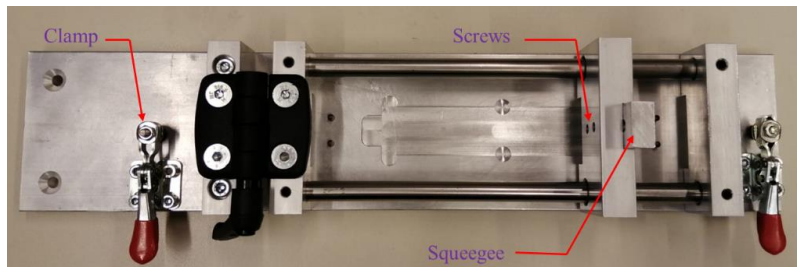

Figure 8. The designed fixture for bonding the elastic strip onto the substrate.

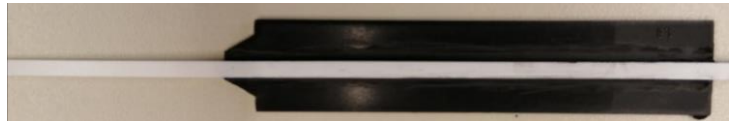

Figure 9. The prepared sample for the peeling test.

### 2.4. Peeling tests

Peeling tests have been conducted according to the ASTM-D1876 standard, employing the Zwick/Roell universal testing machine shown in Figure 10. To perform peeling, the specimen should be inserted between the clamp and base body of the peel kit, fixing the free strip within the grip installed on the moving crosshead. The peeling force has been recorded by a loading cell mounted on the fixed crosshead, while the peel extension has been measured based on the absolute crosshead travel. Moreover, the control parameters such as the loading rate, preloading, time save interval, etc. for performing the test have been selected after preliminary tests and specified thanks to the testXpert II V 3.41 software interface. As mentioned before, six different substrates in terms of roughness have been considered, i.e. those created through steel blocks with $R_z$=0.6, 0.7, 1.8, 3.0, 7.5 μm. Moreover, smooth substrates have been tested as well. Furthermore, specimens with adhesive thicknesses of t= 0.5, 0.8, 1.2, 1.9 and 3.0 mm have been prepared, to elucidate the effect of adhesive thickness on the 90-degree peeling response of the samples. For each test, at least three samples have been tested for a total of 150 peeling acquisitions.

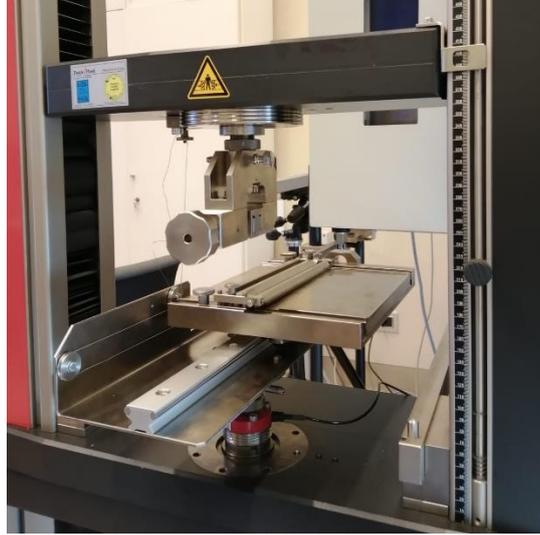

*Figure 10. Peeling test setup.*

## 3. Results and discussion

The results corresponding to the conducted peeling tests are elaborated in this section.

### 3.1. Peeling response

For a linearly elastic and inextensible strip and a 90-degree peeling test, the adhesive fracture energy ($G$) can be evaluated through the ratio between the peeling force ($F$) and the strip width ($w$), i.e. as $\frac{F}{w}$ according to Rivlin equation [26]:

$$\frac{F}{w} = \frac{G}{1 - \cos\theta} \tag{1}$$

where $\theta$ is the peeling angle.

Hence, the value corresponding to the plateau of the $\frac{F}{w}$ - $\Delta$ curve, where $\Delta$ is the peel extension, represents the adhesive fracture energy, see Figure 11. It should be noted that, hereafter, in the presence of roughness, the last 25 mm of peel extension represent the corresponding peeling response of the rough part.

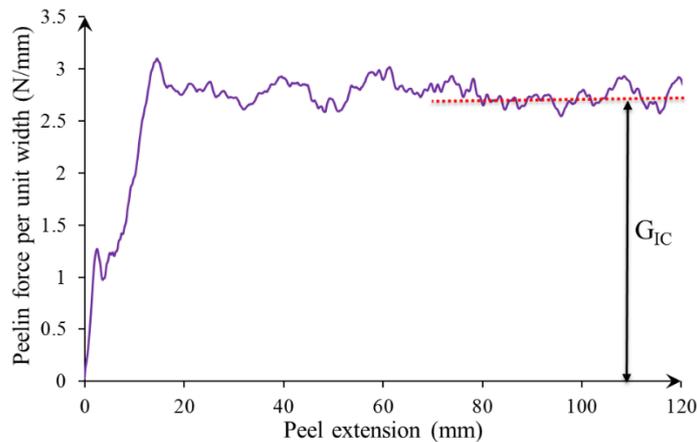

*Figure 11. Peeling force per unit width vs. peel extension.*

## 3.2. Loading rate (LR) effect

According to the ASTM-D1876 standard, it is recommended to apply the displacement at a constant head speed of 254 mm (10 in.) per minute. In our tests, three crosshead speed values (80, 254, and 1000 mm/min) have been considered to investigate the loading rate effect on the peeling response of the samples. Key results are shown in Figure 12. By increasing the crosshead speed, there is no noticeable effect on the peeling response of the specimens, particularly on the fracture energy value. However, higher loading rates avoid high-frequency oscillations in the measured peeling force values, smoothing out the curve.

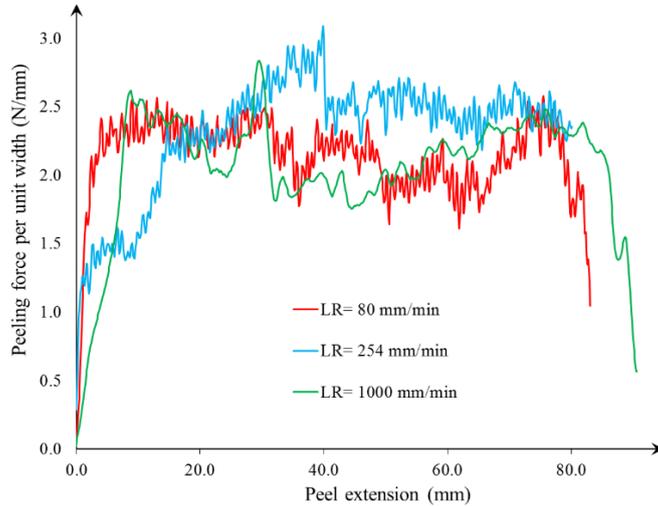

*Figure 12. Effect of loading rate on the peeling response of the specimens.*

## 3.3. Adhesive thickness effect

As mentioned in Sec. 2.4, adhesive thicknesses $t=$ 0.5, 0.8, 1.2, 1.9 and 3.0 mm have been investigated to assess the impact of this parameter on the peeling response of the adhesive joint. Figure 13 illustrates the peeling force per unit width of the strip vs. the peel extension. Such results correspond to smooth substrates. The adhesive thickness plays a significant role in the peeling response and the fracture energy varies from 1.7 to 3.4 N/mm by increasing the adhesive thickness from 0.5 to 3.0 mm. In other words, the higher the adhesive thickness, the higher the fracture energy (except for $t=1.2$ mm and 1.9 mm which displayed similar fracture energies).

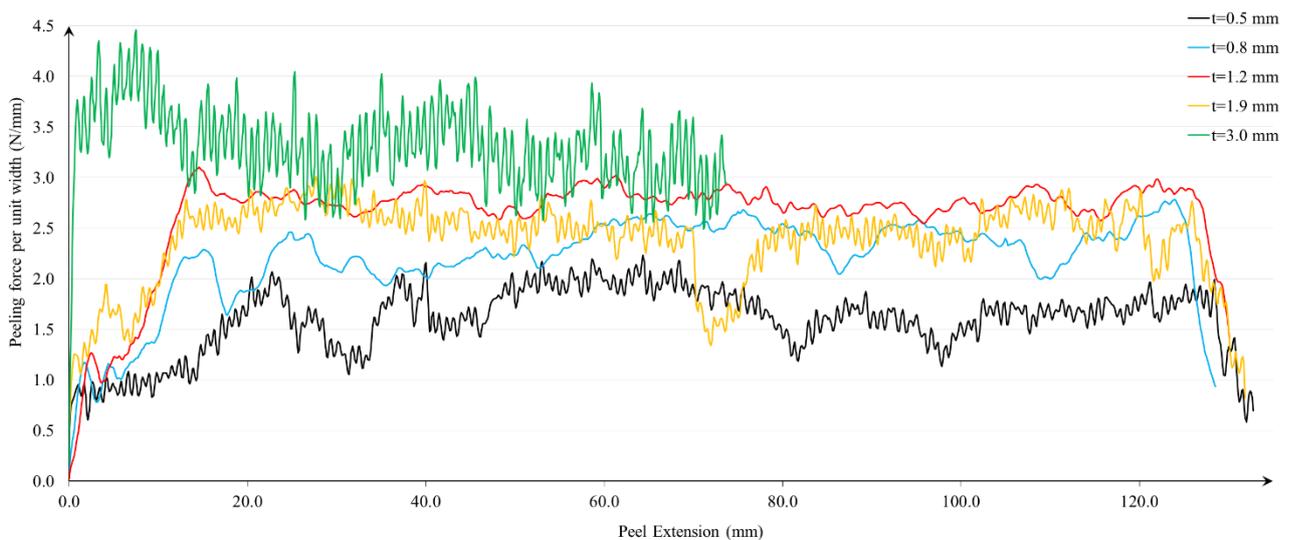

*Figure 13. The adhesive thickness effect on the peeling response of the adhesive joint.*

## 3.4. Roughness effect

In the present subsection, the influence of the substrate roughness on the peeling response has been investigated. Results from the finest to the coarsest substrate for different adhesive thicknesses are plotted in Figures 13 – 17. A closer look at these curves reveals that, in some cases, depending upon the adhesive thickness and the r.m.s. roughness elevation, roughness can affect the peeling force, while in other cases its effect is negligible. Thus, by introducing a dimensionless parameter given by the adhesive thickness to r.m.s. roughness ratio $\left(\alpha = \frac{t}{\sigma}\right)$, we can propose an interpretation. Here, $\sigma$ denotes the root mean square elevation of the representative rough surface, which has been elaborated using methods presented in Sec. 2.2. According to the results collected in Figures 13 - 17, Table 3 classifies the peeling response of the specimens with different adhesive thickness affected by substrate roughness based on the parameter $\alpha$. Evidently, there is a critical value for adhesive thickness to r.m.s. roughness ratio ($\alpha = 246$) such that for configurations having α<246, the substrate roughness does matter to the adhesive peeling response. On the other hand, for thickness to roughness ratios larger than 246, the substrate roughness effect becomes negligible.

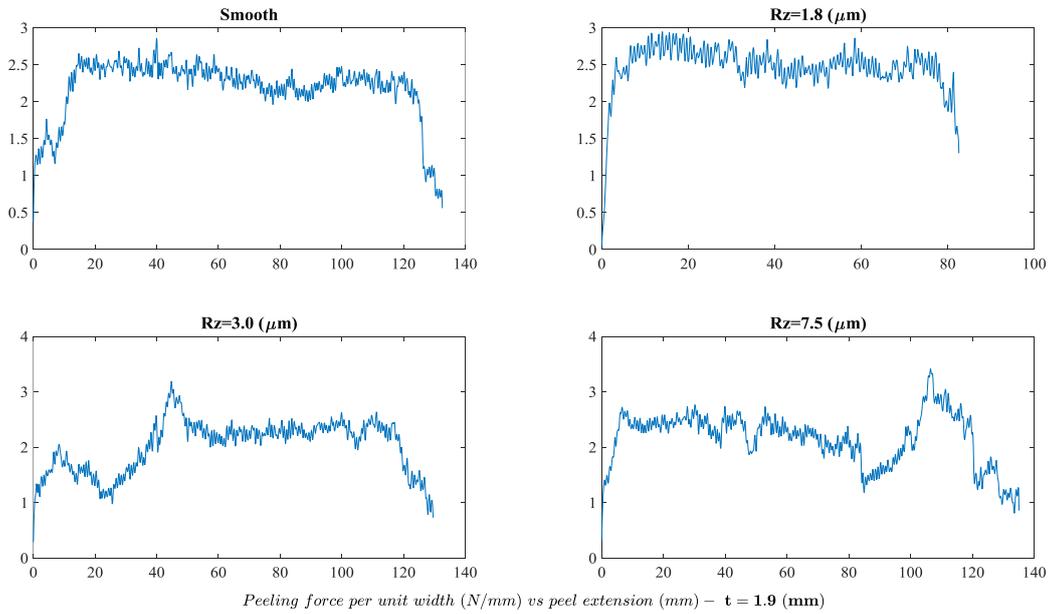

*Peeling force per unit width (N/mm) vs peel extension (mm) – $\mathbf{t = 1.9}$ (mm)*

*Figure 14. Peeling response of the substrate with different roughnesses and adhesive thickness of t=1.9 mm.*

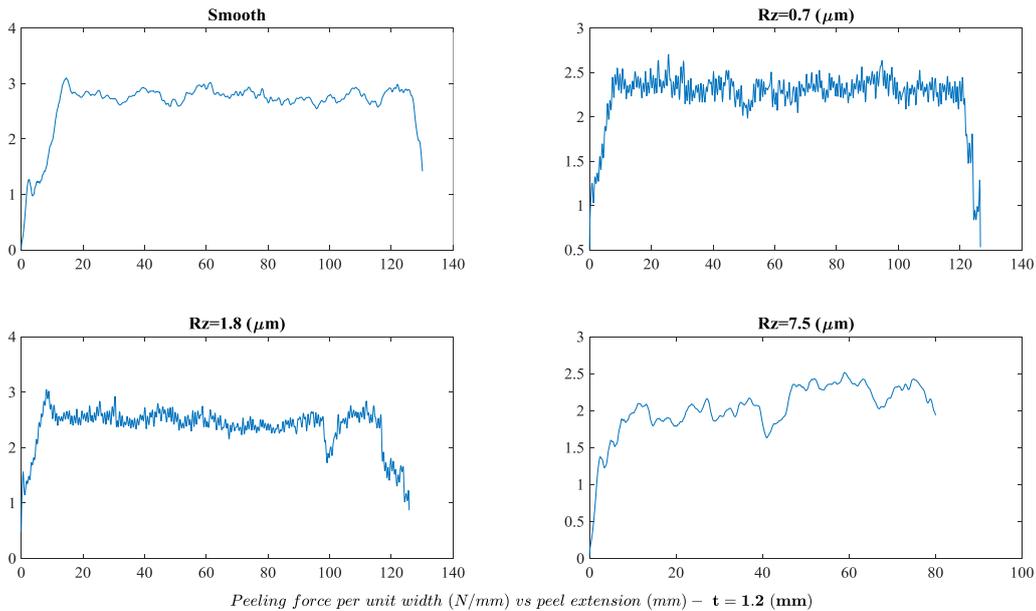

*Peeling force per unit width (N/mm) vs peel extension (mm) – $\mathbf{t = 1.2}$ (mm)*

*Figure 15. Peeling response of the substrate with different roughnesses and adhesive thickness of t=1.2 mm.*

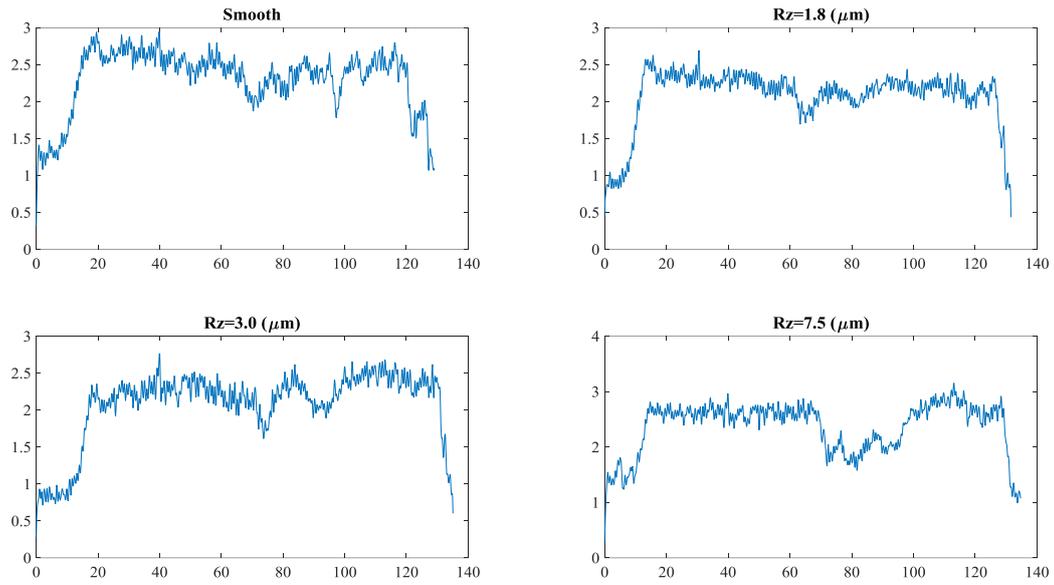

*Figure 16. Peeling response of the substrate with different roughnesses and adhesive thickness of t=0.8 mm.*

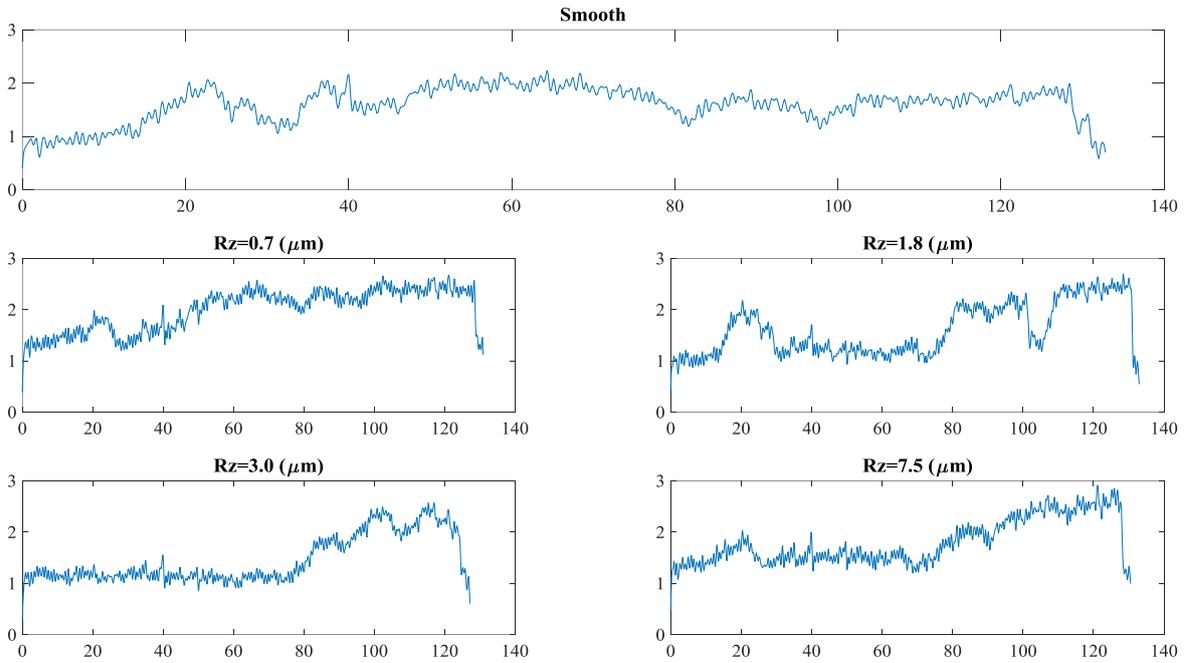

*Figure 17. Peeling response of the substrate with different roughnesses and adhesive thickness of t=0.5 mm.*

*Table 3. Dimensionless adhesive thickness to roughness ratio for all scenarios.*

| $R_z$ (μm) | $\sigma$ (μm) | α | | | | |
|---|---|---|---|---|---|---|
| | | t=0.5 (mm) | t=0.8 (mm) | t=1.2 (mm) | t=1.9 (mm) | t=3.0 (mm) |
| 0.6 | 1.14 | 439 | 702 | 1053 | 1667 | 2632 |
| 0.7 | 2.20 | 227 | 363 | 545 | 864 | 1364 |
| 1.8 | 3.25 | 154 | 246 | 369 | 585 | 923 |
| 3.0 | 4.03 | 124 | 199 | 298 | 471 | 744 |
| 7.5 | 11.36 | 44 | 70 | 106 | 167 | 264 |

## 4. Conclusion

In the present contribution, the impact of surface roughness on the mechanical response of adhesive joints has been addressed. To this aim, an experimental protocol has been proposed to characterize the Silicone-based adhesive response, which has been bonded onto the substrates including a rough part, through peeling tests. A statistically representative rough surface was determined using the normalized distribution of the height filed and Dynamic Space Warping (DSW) analysis of acquired rough surfaces employing a confocal profilometer. For samples' preparation, a specific fixture was designed to bond a flexible strip onto the substrate capable of controlling the adhesive thickness and then the peeling tests were conducted following the ASTM-D1876 norm. Based on the experimental results, the following statements have been achieved:

- The loading rate has a negligible effect within the considered range of rates on the adhesion energy.
- An increase in the adhesive thickness leads to higher adhesion energy for smooth surfaces.
- The dimensionless parameter ($\alpha$), defined as the ratio between the adhesive thicknesses and the root mean square roughness, can be used as an indicator for evaluating the importance of roughness on the peeling response. In general, it has been found that roughness has an impact for $\alpha<246$.